\documentclass[prl,floatfix,reprint,amsmath,amssymb,longbibliography]{revtex4-2}

\usepackage[locale=US]{siunitx}
\DeclareSIUnit\angstrom{\text {Å}}
\usepackage{orcidlink}
\usepackage{textcomp}
\usepackage[capitalise]{cleveref}
\usepackage[]{tikz}
\usetikzlibrary{decorations.pathreplacing,calligraphy,arrows.meta,calc,math,decorations.text}
\tikzset{>=latex}
\usepackage{mathptmx}
\usepackage{pgfplots}
\pgfplotsset{width=3in,compat=1.9}

\usepackage{hyperref}

\definecolor{fzjblue}{HTML}{023d6b}
\definecolor{fzjlightblue}{HTML}{adbde3}
\definecolor{fzjgrey}{HTML}{ebebeb}
\definecolor{fzjgreen}{HTML}{b9d25f}
\definecolor{fzjyellow}{HTML}{faeb5a}
\definecolor{fzjviolet}{HTML}{af82b9}
\definecolor{fzjred}{HTML}{eb5f73}
\definecolor{fzjorange}{HTML}{fab45a}

\definecolor{fzjlightyellow}{HTML}{fef9ce}

\definecolor{dxa_bcc_green}{HTML}{00ff00}
\definecolor{dxa_bcc_pink}{HTML}{ff4dcc}
\definecolor{dxa_bcc_blue}{HTML}{3380ff}
\definecolor{dxa_bcc_red}{HTML}{e63333}

\begin{document}

\title{Conservative adaptive-precision interatomic potentials}

\author{David Immel\,\orcidlink{0000-0001-5143-8043}}
\affiliation{Jülich Supercomputing Centre (JSC), Forschungszentrum Jülich, Jülich, Germany}

\author{Ralf Drautz\,\orcidlink{0000-0001-7101-8804}}
\affiliation{Interdisciplinary Centre for Advanced Materials Simulations (ICAMS), Ruhr Universität Bochum, Bochum, Germany}

\author{Godehard Sutmann\,\orcidlink{0000-0002-9004-604X}}
\email[]{g.sutmann@fz-juelich.de}
\affiliation{Jülich Supercomputing Centre (JSC), Forschungszentrum Jülich, Jülich, Germany\\Interdisciplinary Centre for Advanced Materials Simulations (ICAMS), Ruhr Universität Bochum, Bochum, Germany}

\date{\today}

\begin{abstract}
Adaptive precision molecular dynamics simulations have developed along energy- and force-coupling approaches, which allow for a continuous transition between different particle descriptions or interaction potentials. Most approaches consider different (fixed) spatial regions, which control the transition between the descriptions and consequently avoid a consistent momentum-conserving Hamiltonian description. We present here a new approach to fully integrate the coupling into a Hamiltonian, therefore allowing for a conservative description, which, by design, guarantees both energy and momentum conservation. By coupling a fast EAM potential to a highly accurate ACE potential, we verify numerically the conservation properties and show that one can achieve -- dependent on both the potential and the atomistic system -- a speedup of one or two orders of magnitude compared to a pure ACE simulation.
\end{abstract}

\maketitle

Computer simulations of atomistic systems are a cornerstone of research in Physics, Chemistry, Biology and Materials Science.
Today atomistic simulations occupy a large part of the worlds supercomputers and high-performance computing clusters~\cite{jupiter,frontier,aurora,anton3,noctua2}.
Increasing computational efficiency of atomistic simulations is therefore an integral part of resource-responsible and energy-efficient computing.

Often the outcome of simulations is determined by only a small number of atoms.
For example, in simulations of crack propagation, bond-breaking at the crack tip is decisive~\cite{PhysRevLett.96.095505}.
In simulations of plastic deformation, the generation and propagation of dislocations are key~\cite{paper_nanoindentation}.
The same holds for defect nucleation sites or reaction centers in biology~\cite{doi:10.1021/acs.jctc.0c01112}.
For many years researchers have developed frameworks to exploit hierarchical structures in simulations, where some atoms are more important or decisive than others~\cite{adaptive_precision_potentials, paper_nanoindentation, birks2025efficientaccuratespatialmixing, ml_ff_coupling_force_based, doi:10.1021/acs.jctc.0c01112, PhysRevLett.96.095505, HAdReS_2013_energy_based, delle_site, qm_mm_coupling_hard_coded_zones1, KERDCHAROEN1996313, soft_matter_polymer_adaptive_precision, AdReS_2008_force_based, soft_matter_polymer_adaptive_precision_3, md_mpc_coupling_hard_coded_zones1, csanyi.g.2004a, DeVita1997, Csanyi_2005, energy_based_switching_qmmm}.
A common strategy is to model the decisive atoms with a method that enables accurate atomic forces and energies at a high computational cost and to model atoms that are less important with a simpler, less accurate description of forces and energies at a smaller numerical cost.
The combination of precise and fast simulation models within one simulation leads to a speed-up compared to a simulation where all atoms are described by the precise model by a factor
\begin{equation}
s = \frac{ t^{(\text{p})}}{ x \,  t^{(\text{p})} + (1-x) \, t^{(\text{f})} + t^{(\text{d})}} \,,
\label{eq:cp:speedup}
\end{equation}
where $t^{(\text{p})}$ and $ t^{(\text{f})} $ are the execution times per atom of the precise and fast models, respectively, and $t^{(\text{d})}$ is an overhead that we will specify later.
In many simulations the fraction of atoms $x$ that need to be described by the precise model is of the order of 1\% or 1\textperthousand~\cite{PhysRevLett.96.095505,paper_nanoindentation,doi:10.1021/acs.jctc.0c01112}.
Often one can assume that the evaluation time for the fast model~\cite{aside_potentials}\nocite{gaussian_approximation_potentials,spectral_neighbor_analysis_potentials,moment_tensor_potentials,ace,JE_Lennard-Jones_1931,PhysRevLett.50.1285,PhysRevB.31.5262,PhysRevB.90.104108,10.1063/12.0000881,NOVOSELOV201946,pace,doi:10.1021/acs.jctc.2c01149,Stark_2024} as well as the overhead~\cite{adaptive_precision_potentials} is orders of magnitude smaller than the computational cost of the precise model, $t^{(\text{f})} \approx t^{(\text{d})}  \ll t^{(\text{p})}$, such that a potential speedup of orders of magnitude can be expected~\cite{adaptive_precision_potentials,paper_nanoindentation}.
The coupling of fast and precise models therefore is very promising to allow for simulations of either much larger system sizes or diminished resource requirements at constant simulation size.

The coupling of two different models in one simulation presents a formidable scientific challenge that has a long history in concurrent multi-scale modeling.
Traditionally regions for precise and fast models are identified before the start of a simulation and the challenge is a sound description of the interface between the two regions~\cite{qm_mm_coupling_hard_coded_zones1, doi:10.1021/acs.jctc.0c01112, ml_ff_coupling_force_based, HAdReS_2013_energy_based, delle_site, md_mpc_coupling_hard_coded_zones1,csanyi.g.2004a,DeVita1997,Csanyi_2005}.

\begin{figure}
\centering
\begin{tikzpicture}[x=1in,y=1in,style={thick}]
\tikzmath{
    \pw = 3.15;
    \cbh = 0.1; \cbw=0.7;
    \dsD = 1.8;
    \rsh = 0.8;
}
\node[coordinate] (rs_sw) at (0,-2) [] {};
\tikzmath{\ph = \pw/2000*825;}
\node[coordinate] (p_h) at (0,\ph) [] {};
\node[coordinate] (p_w) at (\pw,0) [] {};
\node[coordinate] (cb_h) at (0,\cbh) [] {};

\node[coordinate] (ds_sw) at ($(rs_sw) + (0,0.4) + (0,\rsh)$) [] {};
\tikzmath{
    \ri = 0.35;
    \ro = 0.6;
    \dOx = 0.41;
    \dOy = 0.47;
}
\node[coordinate] (d0) at ($\dsD*(\dOx, \dOy) + (ds_sw)$) [] {};
\node[coordinate] (ds_w) at (\dsD,0) [] {};
\node[coordinate] (ds_h) at (0,\dsD) [] {};
\node[coordinate] (ds_c) at ($(ds_sw) + 0.5*(ds_w) + 0.5*(ds_h)$) [] {};
\node[coordinate] (ds_ne) at ($(ds_sw) + (ds_w) + (ds_h)$) [] {};
\draw[] ($(ds_sw)-(0.25,0)+(ds_h)$) node[anchor=south east] {a)};
\tikzmath{\dsDhalf = 0.48*\dsD; }
\tikzmath{\rtot = (sqrt((\dOx-0.5)^2 + (\dOy-0.5)^2) + 0.5) * \dsD;}
\begin{scope}
\clip (ds_c) circle (\dsDhalf);
\draw[inner color=fzjgreen,outer color=teal, draw=none] (d0) circle (\rtot);
\end{scope}
\draw[->] (ds_sw) -- ++ (ds_w) node[anchor=north] {$d_\text{x}$};
\draw[->] (ds_sw) -- ++ (ds_h) node[anchor=east]  {$d_\text{y}$};
\draw[dashed] ($(ds_sw) + \dOx*(ds_w)$) node[anchor=north] {$d_\text{x}^{(0)}$} -- (d0) -- ($(ds_sw) + \dOy*(ds_h)$) node[anchor=east] {$d_\text{y}^{(0)}$};
\draw[white] (d0) node[anchor=south] {fast model} node[anchor=south, yshift=2ex] {use};
\tikzmath{\ritot = \ri*\dsDhalf;}
\draw[white] (d0) circle (\ritot);
\tikzmath{\rotot = \ro*\dsDhalf;}
\draw[white] (d0) circle (\rotot);
\tikzmath{\tmpr = (\ri+\ro)/2*\dsDhalf;}
\path[postaction={decorate,decoration={raise=-0.5ex, text along path, text color=white, text align=center, reverse path, text={transition region}}}] ($(d0)+(0:\tmpr)$) arc (0:180:\tmpr);
\tikzmath{\tmpr = (\ro)*\dsDhalf;}
\path[postaction={decorate,decoration={raise=2ex, text along path, text color=white, text align=center, reverse path, text={precise model required}}}] ($(d0)+(0:\tmpr)$) arc (0:180:\tmpr);
\tikzmath{\tmpr = 1.2*\ro*\dsDhalf;}
\draw[color=white,->] (d0) -- ++(-50:\tmpr) node [pos=1.0,below] {$\pmb{d}_i$};
\node[coordinate] (cb1_nw) at ($(ds_ne) + (0.1,-0.2)$) [] {};
\node[coordinate] (cb2_sw) at ($(ds_ne) + (0.1,+0.2) - (ds_h)$) [] {};
\node[coordinate] (cb1_sw) at ($(cb1_nw) - (cb_h)$) [] {};
\node[coordinate] (cb2_nw) at ($(cb2_sw) + (cb_h)$) [] {};
\node[coordinate] (cb_w) at (\cbw,0) [] {};
\draw[] (ds_ne) node[anchor=south west] {b)};
\draw[left color=fzjgreen,right color=teal] (cb1_sw) rectangle ($(cb1_sw) + (cb_h) + (cb_w)$);
\draw[] ($(cb1_sw) + (cb_w) + 0.5*(cb_h)$) node[anchor=west] {$d_i$};
\node[anchor=south west,inner sep=0] at (cb2_sw) {\frame{\includegraphics[width=\cbw in,height=\cbh in,angle=180]{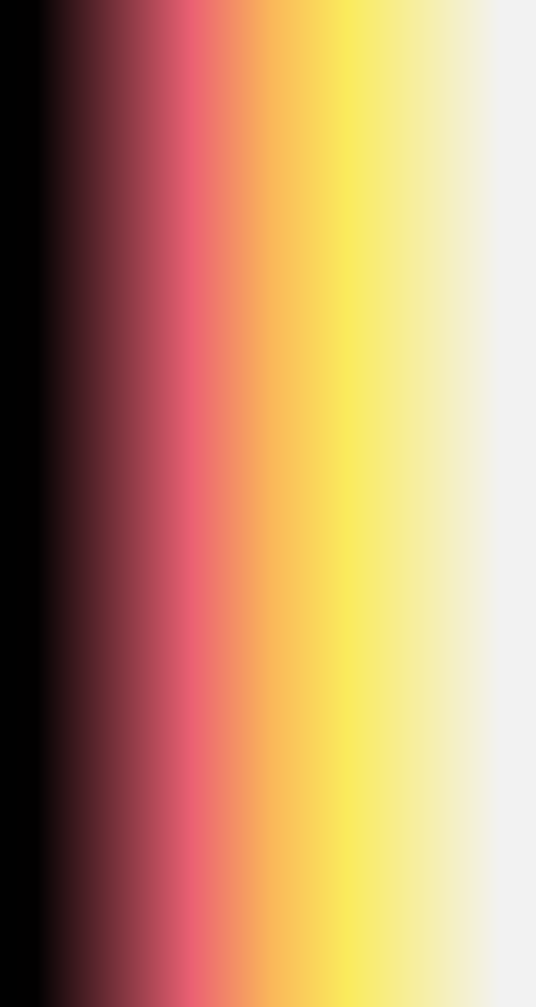}}};
\draw[] ($(cb2_sw) + (cb_w) + 0.5*(cb_h)$) node[anchor=west] {$\lambda_i$};
\draw[] ($(cb2_sw) + 0.5*(cb_w) + 1*(cb_h)$) node[anchor=south] {\rotatebox{0}{\scriptsize transition range}};
\draw[->] ($(cb1_sw) + (cb_w) + (0.1,-0.05)$) to[out=-90,in=90] node[midway,anchor=west] {\rotatebox{270}{\scriptsize \cref{eq:cp:lambda}}} ($(cb2_sw) + (cb_h) + (cb_w) + (0.1,+0.05)$);
\tikzmath{\cblo = \ritot / \rtot * \cbw;}
\tikzmath{\cbhi = \rotot / \rtot * \cbw;}
\draw[thick] ($(cb1_sw) + 1.2*(cb_h) + (\cblo,0)$) node[anchor=south] {\scriptsize$d_\text{lo}$} -- ++ ($-1.0*(cb_h)$) -- ($(cb2_sw) + (cb_h)$) node[midway,sloped,above,xshift=-0.3ex] {\scriptsize fast} -- ++ ($-1.2*(cb_h)$) node[anchor=north] {$1$};
\draw[thick] ($(cb1_sw) + 1.2*(cb_h) + (\cbhi,0)$) node[anchor=south] {\scriptsize$d_\text{hi}$} -- ++ ($-1.0*(cb_h)$) -- ($(cb2_sw) + (cb_h) + 1.01*(cb_w)$) node[midway,sloped,above,xshift=-0.2ex] {\scriptsize precise} -- ++ ($-1.2*(cb_h)$) node[anchor=north] {$0$};
\draw[thick,white] ($(cb1_sw) + (\cblo,0)$) -- ++ (cb_h) ($(cb1_sw) + (\cbhi,0)$) -- ++ (cb_h);
\node[coordinate] (rs_w) at (3.0,0) [] {};
\node[coordinate] (rs_h) at (0,\rsh) [] {};
\node[coordinate] (tip) at ($(rs_sw) + 0.5*(rs_w) + 0.5*(rs_h)$) [] {};
\fill[white!95!black] (tip) -- ($(rs_sw) + 0.05*(rs_h)$) -- (rs_sw) -- ++ ($0.9*(rs_w)$) -- ++ (rs_h) -- ($(rs_sw) + (rs_h)$) -- ++ ($-0.05*(rs_h)$) -- cycle;
\begin{scope}
\clip ($(rs_sw) + 0.5*(rs_h)$) rectangle ++ ($0.51*(rs_w) + 0.5*(rs_h)$);
\node[anchor=south east] at ($(tip) + (0,-0.2) + 0.33*(rs_w)$) {\includegraphics[height=3.0in,width=0.3in,angle=80]{cb_myhot.png}};
\end{scope}
\begin{scope}
\clip (rs_sw) rectangle ++ ($0.51*(rs_w) + 0.5*(rs_h)$);
\node[anchor=north east] at ($(tip) + (0,+0.2) + 0.33*(rs_w)$) {\includegraphics[height=3.0in,width=0.3in,angle=100]{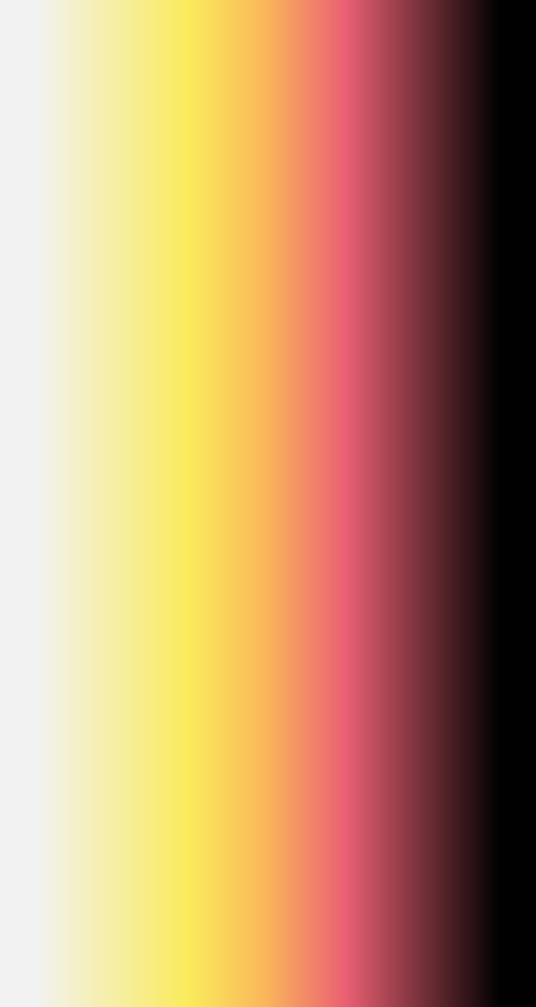}};
\end{scope}
\tikzset{
  cb shading/.code args={}{
    \pgfdeclareradialshading{cb}{\pgfpoint{0cm}{0cm}}%
    {
      color(0pt)=(black);
      color(0.07cm)=(black);
      color(0.32cm)=(fzjred);
      color(0.45cm)=(fzjorange);
      color(0.56cm)=(fzjyellow);
      color(0.77cm)=(white!95!black);
      color(0.8818cm)=(white!95!black)
    }
    \pgfkeysalso{/tikz/shading=cb}
  },
}
\begin{scope}
\clip ($(rs_sw) + 0.5*(rs_w)$) rectangle ++ ($0.5*(rs_w) + 1.0*(rs_h)$);
\shade[cb shading={}] (tip) circle (0.31);
\end{scope}
\begin{scope}
\clip ($(rs_sw) + 0.5*(rs_h) - (0,0.270)$) -- ++(-40:0.4) -- ($(rs_sw) + (0.3,0)$) -- (rs_sw) -- cycle;
\node[anchor=south west] at ($(rs_sw) -(0.05,0.05)$) {\includegraphics[width=0.30in,angle=180]{cb_myhot_flipped.png}};
\end{scope}
\begin{scope}
\clip ($(rs_sw) + 0.5*(rs_h) + (0,0.280)$) -- ++(40:0.4) -- ($(rs_sw) + (rs_h) + (0.3,0)$) -- ($(rs_sw) + (rs_h)$) -- cycle;
\node[anchor=north west] at ($(rs_sw) + (rs_h) +(-0.05,0.05)$) {\includegraphics[width=0.30in,angle=180]{cb_myhot_flipped.png}};
\end{scope}
\node[coordinate] (cs_rs) at ($(rs_sw) - (0.05,0.05)$)  [] {};
\draw[->] (cs_rs) -- ++ ($0.9*(rs_w)$) -- ++ (0.1,0) node[anchor=north] {$x$};
\draw[->] (cs_rs) -- ++ (rs_h) -- ++ (0,0.1) node[anchor=east]  {$y$};
\draw (cs_rs) ++ (rs_h) ++ (-0.2,0) node[anchor=south east] {c)};
\draw ($(rs_sw) + 0.0*(rs_w) + 0.5*(rs_h)$) node[anchor=west] {precise model} ++ ($0.9*(rs_w)$) node[anchor=east] {fast model};
\end{tikzpicture}
\caption{\label{fig:cp:vis:scheme}
The deviation $\pmb{d}_i$ (a) is measured in the descriptor space and used to calculate (b) the switching parameter $\lambda_i$ (c) between fast and precise model for all atoms $i$.
}
\end{figure}

However, this requires knowledge of expected key atomistic events before the start of the simulation, while for many simulations it is desirable that regions of fast or precise description appear or vanish dynamically as dictated by the trajectories of the atoms.
Variable switching between fast and precise models can be achieved with the help of descriptors that quantify the atomic environment of the atoms.
For each atom $i$ the descriptor $\pmb{d}_i$ is evaluated.
The fast model is trained as a surrogate model to reproduce the precise model in the vicinity of reference descriptor values $\pmb{d}^{(0)}$.
For example, the reference descriptor $\pmb{d}^{(0)}$ corresponds to a perfect lattice at equilibrium volume.
Then for small deviations of $\pmb{d}$ from $\pmb{d}^{(0)}$, the fast potential provides atomic forces and energies that closely match the precise model.
At larger deviations of $\pmb{d}$ from $\pmb{d}^{(0)}$, the fast surrogate model is insufficient and the precise model is used as illustrated in \cref{fig:cp:vis:scheme}a.

The transition between fast and precise model is achieved by a smooth switching function $\lambda(\pmb{d})$ (cf. \cref{fig:cp:vis:scheme}b), i.e.,
\begin{equation}
\lambda_i = f^{(\text{trans})}\left(d_i, d_\text{lo}, d_\text{hi}\right) \,,
\label{eq:cp:lambda}
\end{equation}
where $f^{(\text{trans})}$ is 1 for $d_i\leq d_\text{lo}$, decays differentiable to 0 until $d_i = d_\text{hi}$ and stays zero for $d_i > d_\text{hi}$ (see Ref. \cite{sm} for equation and plot of $f^{(\text{trans})}$).

There are different ways to combine fast and precise models.
An intuitive approach consists in superimposing the atomic forces of two potentials weighted by a switching parameter $\lambda_i$ evaluated for atom $i$ as~\cite{PhysRevLett.96.095505, ml_ff_coupling_force_based, KERDCHAROEN1996313}
\begin{equation}
\pmb{F}_i= \lambda_i \pmb{F}_i^\text{(f)} + (1-\lambda_i) \pmb{F}_i^\text{(p)}\,.
\label{eq:cp:force:mix}
\end{equation}
However, the energy that corresponds to the force is not known and, consequently, force mixing simulations do not conserve energy and require a thermostat to prevent spontaneous heating during a simulation; besides, \cref{eq:cp:force:mix} violates Newton's third law, i.e., it does not conserve the momentum either~\cite{birks2025efficientaccuratespatialmixing}.
A conservative potential may be obtained by mixing the atomic energies directly,~\cite{adaptive_precision_potentials}
\begin{equation}
{E}_i= \lambda_i {E}_i^\text{(f)} + (1-\lambda_i) {E}_i^\text{(p)}\,.
\label{eq:cp:energy:hyb}
\end{equation}
Thereby, the Hamiltonian is in the absence of external fields given as $H=\sum_i(E_i + T_i)$, where $T_i$ denotes the kinetic energy of an atom $i$.

The difficulty -- caused by superimposing the energies -- is that the forces on the atoms,~\cite{adaptive_precision_potentials}
\begin{equation}
\begin{split}
\pmb{F}_i  = -\nabla_i \sum_k E_k = \sum_k \big(&- \lambda_k \nabla_i E_k^\text{(f)} - (1 - \lambda_k) \nabla_i E_k^\text{(p)} \\
&+ (\nabla_i\lambda_k) (E_k^\text{(p)} - E_k^\text{(f)})\big)\,,
\end{split}
\label{eq:cp:force:hyb}
\end{equation}
may deviate significantly from the forces of the precise reference potential: If the energy of the fast and precise potential differ in the switching region, the gradient of the switching parameter contributes to the force, and a steep gradient of the switching parameter can induce large forces neither present in the precise reference nor in the fast potential (cf. Appendix A).
Furthermore, $\pmb{F}_i$ (\cref{eq:cp:force:hyb}) depends on the switching parameters $\lambda_k$ of neighboring atoms $k$.
Therefore, a smooth transition is required between both the reference and the fast potential to minimize artifacts in forces from the coupling region.

A simple and common way to achieve such a transition between precise and fast model is to interpolate the switching parameter dependent on an atom's position from 1 to 0 between spatial reference positions~\cite{HAdReS_2013_energy_based, delle_site, ml_ff_coupling_force_based, md_mpc_coupling_hard_coded_zones1}, which implies that the force (cf. \cref{eq:cp:force:hyb}) caused by the differentiation of the switching function also depends on an atom's position and, therefore, does not conserve the momentum~\cite{aside_momentumconservation,HAdReS_2013_energy_based}.
This example illustrates the challenge of the switching-parameter calculation for energy-based approaches, that we address in this letter.

Therefore, the descriptor $\pmb{d}$, which we use to calculate the switching parameter (cf. \cref{eq:cp:lambda}), cannot depend on spatial constants, but only on atomic distances $\pmb{r}^\text{d}$ and non-spatial constants $\pmb{\beta}$.
In general atomic quantities $\pmb{g}_k(\{\pmb{r}^\text{d}\},\pmb{\beta}$) can be used.
$\pmb{d}_i$ could be calculated from the strain energy to study cracking~\cite{PhysRevLett.96.095505}, depend on the chemical species~\cite{doi:10.1021/acs.jctc.0c01112,aside_zdetection}\nocite{sm}, or depend on multiple quantities as visualized in \cref{fig:cp:vis:scheme}a.
Note that, $d_i$ -- and thus also $\pmb{g}_k$ -- needs to be differentiable as the switching parameter is differentiated for the force calculation (cf. \cref{eq:cp:force:hyb}).

The most direct way to calculate the descriptor $\pmb{d}$ would be $\pmb{d}_i(\pmb{g}_i)$, which would result in accurate energies $E_i^\text{AP}$ (cf. \cref{eq:cp:energy:hyb}).
However, the usage of $\pmb{d}_i(\pmb{g}_i)$ neglects the previously discussed dependency of $\pmb{F}_i^\text{AP}$ on $\lambda_k$ of neighboring atoms $k$ that might have a small $\pmb{d}_k(\pmb{g}_k)$, which could cause an inaccurate $\pmb{F}_i^\text{AP}$.
To prevent this issue, one can instead use a local average of $\pmb{g}_k$, i.e.,
\begin{equation}
\pmb{d}_i = \sum_k \pmb{g}_k(\{\pmb{r}^\text{d}\},\pmb{\beta}) \frac{w(r_{ik})} {\overline{w}_i(\{\pmb{r}^\text{d}\})} - \pmb{d}^{(0)}\,,
\label{eq:cp:descriptor}
\end{equation}
where $w(r)$ is a weighting function, that is zero beyond the cutoff distance $r_\text{w}^\text{cut}$ and $\overline{w}_i(\{\pmb{r}^\text{d}\})=\sum_k w(r_{ik})$ is a normalization factor that ensures the independence of $d_i$ from the number of neighboring atoms within the cutoff distance.

$r_\text{w}^\text{cut}$ should be in the order of magnitude of the cutoff radius of the precise potential, so that atoms with a large $\pmb{g}_k$ increase $\pmb{d}_k$ for all atoms $k$ in their force cutoff.
Thereby, the neighboring switching parameters $\lambda_k(d_k)=0$ can ensure a precise force $\pmb{F}_i$ (cf. \cref{eq:cp:force:hyb}).
Significantly larger averaging radii offer no further advantage as more and more atoms would require the expensive calculation of the precise potential.

The local summation in \cref{eq:cp:descriptor} averages out some fluctuations of $\pmb{g}_i$, that may be caused by the thermal fluctuation of atoms, like the non-differentiable time average does in Ref. \cite{adaptive_precision_potentials}.
Note, that one can modify this approach and include fast/precise atoms to speedup a simulation further analogously to Ref. \cite{paper_nanoindentation} by using a constant time independent switching parameter ($\lambda=0\text{ or }1$) for a subset of atoms without loosing the conservativity (cf. Appendix B).

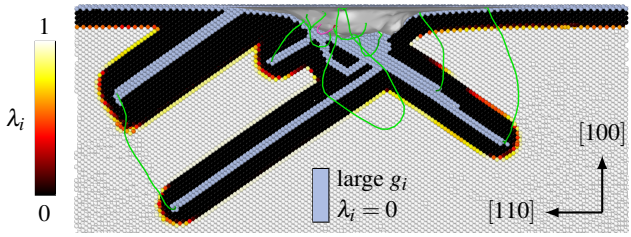
\begin{figure}
\centering
\begin{tikzpicture}[x=1in,y=1in,style={thick}]
\tikzmath{
    \pw = 2.9;
    \ph = \pw/2000*825;
    \cbh = 0.8; \cbw = 0.1;
}
\node[coordinate] (p_w) at (\pw,0) [] {};
\node[coordinate] (p_h) at (0,\ph) [] {};
\node[coordinate] (p0_sw) at (0.0,0) [] {};
\node[coordinate] (p0_ne) at ($(p0_sw) + (p_w) + (p_h)$) [] {};
\node[anchor=south west,inner sep=0] at (p0_sw) {\includegraphics[width=\pw in]{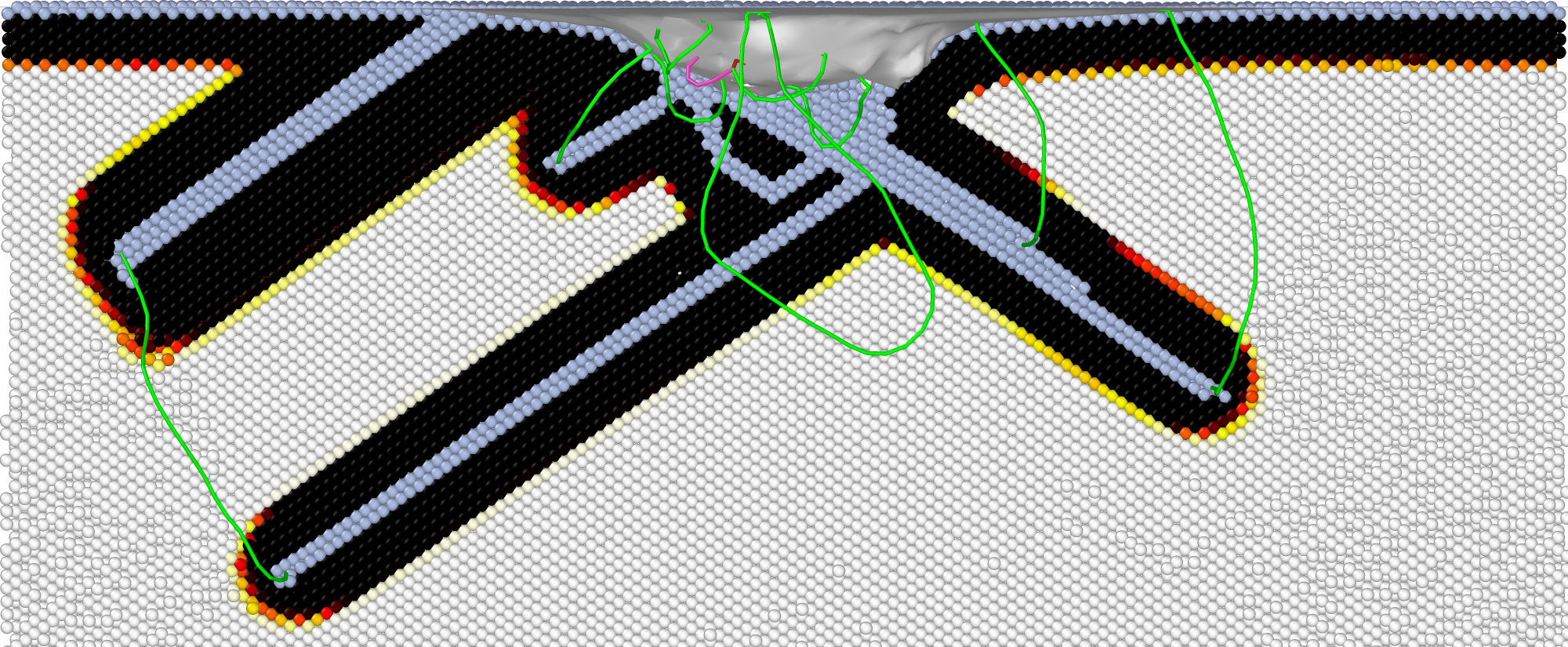}};
\draw ($(p0_sw) + 0.46*(p_w) + (0,0.2)$) node[anchor=east,xshift=0.4em]{\frame{\textcolor{fzjlightblue}{\rule{0.2cm}{0.7cm}}}} node[anchor=south west, yshift=-0.2em] {large $g_i$} node[anchor=north west, yshift=0.2em] {$\lambda_i=0$};
\node[coordinate] (tripod_origin) at ($(p0_sw) + (p_w) + (-0.13,0.11)$) [] {};
\draw[->] (tripod_origin) -- ($(tripod_origin) - (0.3,0)$) node[anchor=east]  {$[110]$};
\draw[->] (tripod_origin) -- ($(tripod_origin) + (0,0.3)$) node[anchor=south] {$[100]$};
\node[coordinate] (p_h) at (0,\ph) [] {};
\node[coordinate] (cb_sw) at ($(-0.2,0) + 0.5*(p_h) - 0.5*(0,\cbh)$) [] {};
\node[anchor=south west,inner sep=0] at (cb_sw) {\frame{\includegraphics[width=\cbh in,height=\cbw in,angle=90]{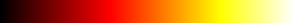}}};
\draw (cb_sw) ++ ($0.5*(\cbw,0)$) node[anchor=north] {$0$} ++ (0,\cbh) node[anchor=south] {$1$};
\draw (cb_sw) ++ ($0.5*(0,\cbh)$) node[anchor=east] {$\lambda_i$};
\end{tikzpicture}
\caption{
Switching parameters $\lambda_i$ -- calculated from $g_i$ via \cref{eq:cp:lambda,eq:cp:descriptor} -- and dislocation lines.
Not shown are the atoms in the lower part of the simulation box, which are calculated with the fast potential, i.e., $\lambda_i=1$.
}
\label{fig:cp:vis:nanoindentation}
\end{figure}

The partially filled d-band of the transition metal tungsten (W) causes directional, i.e., angular-dependent, bonding~\cite{pettifor1995bonding}, which cannot be modeled with central-force potentials~\cite{mrovec2007bond,Marinopoulos01111995,Groeger01122009}.
Thus, W benefits from an adaptive-precision description~\cite{paper_nanoindentation} and serves as test material in the following.
We construct a conservative adaptive precision potential from the adaptive-precision EAM-ACE W potential of Ref. \cite{paper_nanoindentation}, for which we use a differentiable centro-symmetry parameter (CSP) as $g_k$ (cf. Appendix B).
The achievable accuracy of such a conservative EAM-ACE potential is investigated for a nanoindentation snapshot of W taken from a simulation of Ref. \cite{paper_nanoindentation}.
The analyzed nanoindentation snapshot is visualized with OVITO~\cite{ovito} in \cref{fig:cp:vis:nanoindentation} for $r_\text{w}^\text{cut}=\SI{16}{\angstrom}$.
The visualization shows that two layers of atoms $i$ have a significant $g_i$ at the surface, whereas nine layers have $\lambda_i=0$ and the tenth layer transfers with $\lambda_i\in(0,1)$ to the quickly calculated atoms with $\lambda_i=1$ in the eleventh layer.
This comparison between $g_i$ and $\lambda_i$ shows that the calculation of the local average according to \cref{eq:cp:descriptor} changes the switching parameter in the force cutoff of atoms with large $g_i$ towards a precise calculation, which is essential for the force accuracy (cf. Appendix A).
The same effect occurs near dislocation lines, that are identified using the dislocation extraction algorithm\cite{ovito_dxa} of OVITO~\cite{ovito}.
Furthermore, slip traces are visible as the gliding dislocations change the neighborhood of atoms and the CSP-opposite neighbor pairs are not updated (as discussed in Appendix B).

\begin{figure}
\begin{tikzpicture}[x=1in,y=1in]
\newcommand{\asize}{\small}
\node[anchor=south west,inner sep=0] at (0,0) {\includegraphics[width=3.37in]{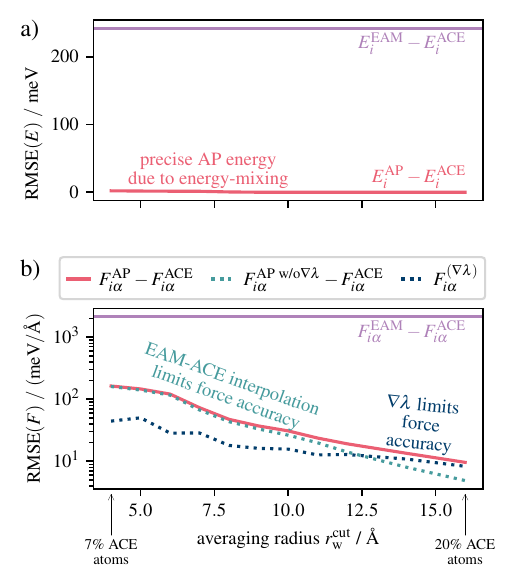}};
\node[coordinate] (inset_sw) at (0.8, 3.0) [] {};
\node[coordinate] (inset_size) at (1,0.503) {} {};
\draw (inset_sw) node[anchor=south west,inner sep=0]
{\includegraphics[width=1.00in]{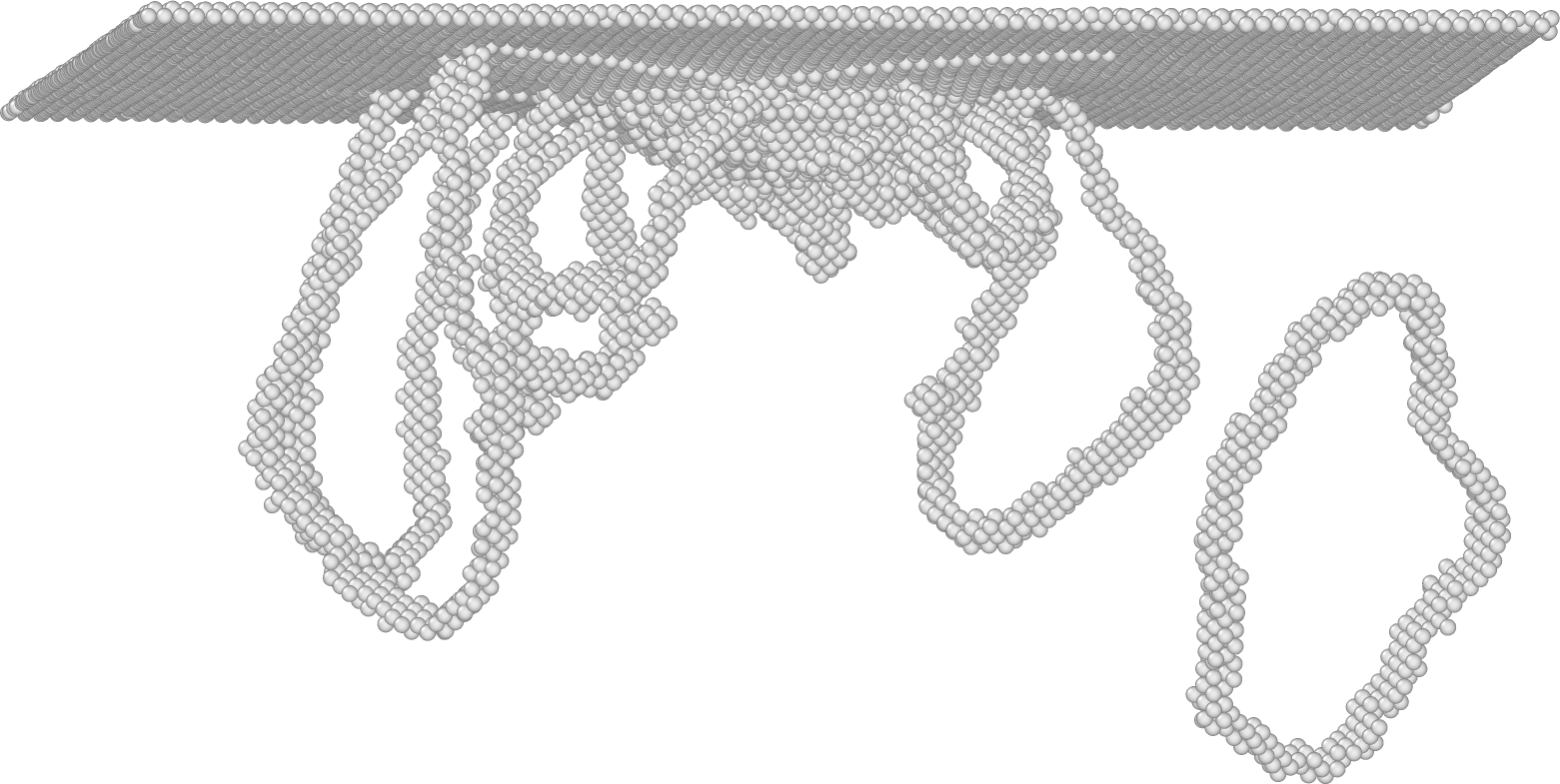}};
\draw[thin] ($(inset_sw) - 0.05*(inset_size)$) node[anchor=south west] {\large c)} rectangle ++ ($1.10*(inset_size)$);
\node[coordinate] (e0) at (2.42,3.45) [] {};
\node[coordinate] (e1) at (2.86,3.45) [] {};
\node[coordinate] (mp) at (2.52,2.85) [] {};
\node[coordinate] (ep) at (2.52,2.68) [] {};
\draw[gray,thick,->] (e0) -- ++(0,-0.6) node[pos=0.5,rotate=270,yshift=1.4ex] {\asize $\lambda_i$} -- (mp) -- (ep);
\draw[gray,thick] (e1) -- ++(0,-0.6) node[pos=0.5,rotate=270,yshift=1.4ex] {\asize $(1-\lambda_i)$} -- (mp);
\draw[color=gray,thick,->] (ep) ++(0,-0.11) -- ++(0,-0.25) -- ++(-1.81,0) node[midway, below,yshift=0.5ex] {\asize $-\nabla_i \sum_k E_k^\text{AP}$} -- ++(0,-0.32);
\end{tikzpicture}
\caption{\label{fig:cp:ap:rmse}
RMSE of the a) potential energies $E_i$ and b) force components $F_{i\alpha}$ of an AP potential compared with the precise ACE potential dependent on the radius $r_\text{w}^\text{cut}$ of the averaging region of the descriptor for the atoms shown in (c) (inset of (a)).
The RMSE of EAM is given for comparison.
}
\end{figure}
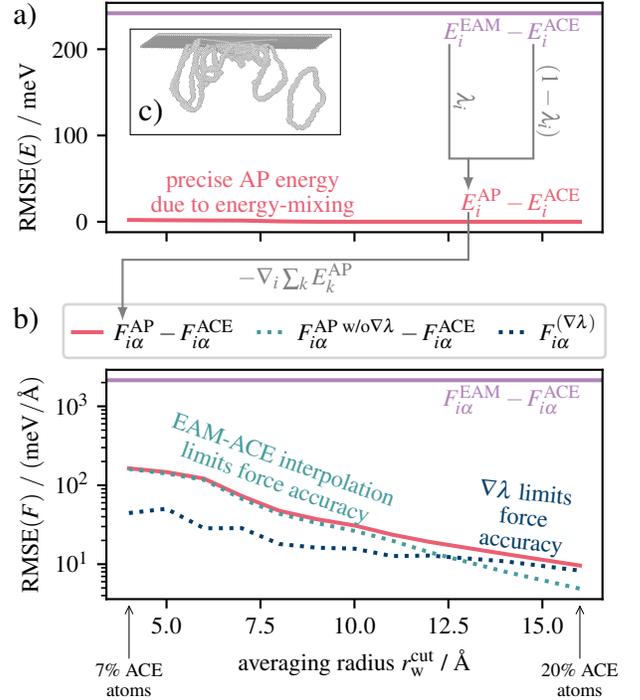

Therefore, to asses the precision of our EAM-ACE potential, we neglect these slip traces by analyzing only the by the non-differentiable CSP identified atoms ($\text{CSP}_i^\text{dyn}>\SI{1.5}{\angstrom^2}$~\cite{paper_nanoindentation}).
Furthermore, we neglect the fixed atoms of the bottom boundary, as they are calculated efficiently as described in Appendix B.
Hence, the in \cref{fig:cp:ap:rmse}c shown group of atoms is analyzed, i.e., atoms near the top surface or near dislocations -- $1.5\%$ of the atoms -- which are not well described by the fast EAM potential.

The root mean square error (RMSE) of AP forces and energies compared with the precise ACE potential is shown in \cref{fig:cp:ap:rmse}.
The precise ACE-energies are used (cf. \cref{fig:cp:ap:rmse}a) due to the energy interpolation with the switching parameter (cf. \cref{eq:cp:energy:hyb}).
The AP forces are conservatively calculated and improve with the averaging radius $r_\text{w}^\text{cut}$ (cf. \cref{fig:cp:ap:rmse}b) up to a RMSE of $\SI{9.6}{\milli\electronvolt/\angstrom}$ at the largest analyzed $r_\text{w}^\text{cut}$ of $\SI{16}{\angstrom}$.
There are two factors, which decrease the force-precision of our AP potentials: 1. force contributions of the fast model and 2. the gradient of the switching parameter (cf. Appendix A).
We investigate the second factor by calculating the force $\pmb{F}_i^{(\nabla\lambda)}$ caused by the differentiation of the switching parameter.
$\pmb{F}_i^{\text{AP\,w/o\,}\nabla\lambda} = \pmb{F}_i^\text{AP} - \pmb{F}_i^{(\nabla\lambda)}$ is a non-conservative AP force independent of the gradient of the switching function and, thus, allows to assess the influence of the fast model on the conservatively calculated AP force.
The force contributions of the fast model dominate for small averaging radii($<\SI{13}{\angstrom}$) but become negligible for larger averaging radii($>\SI{13}{\angstrom}$), for which the gradient of the switching function limits the force-accuracy as shown in \cref{fig:cp:ap:rmse}b.
As the influence of these two mechanisms decays with the averaging radius $r_\text{w}^\text{cut}$, one can adjust the force accuracy at the cost of ACE computations dependent on the numerical requirements.
\begin{figure}
    \centering
    \includegraphics[width=3.37in]{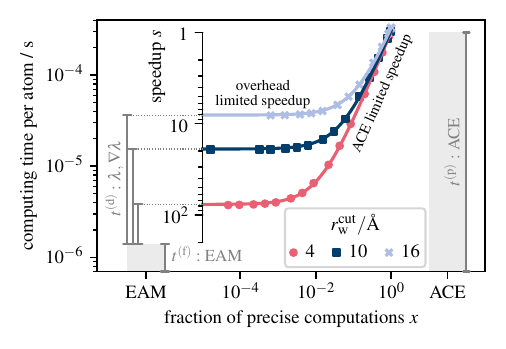}
    \caption{Achievable speedup factor $s$ (cf. \cref{eq:cp:speedup}) and computing time~\cite{aside_cpu} per atom dependent on the fraction $x$ of ACE atoms and the averaging radius $r^\text{cut}_\text{w}$ of the AP potential.}
    \label{fig:cp:speedup}
\end{figure}

The possible speedup $s$ of an AP potential compared to a complete precise simulation depends on the fraction $x$ of precise computations (cf. \cref{eq:cp:speedup}).
We measure this dependency of $s$ on $x$ for the AP EAM-ACE W potential by generating more and more vacancy-interstitial pairs -- which require the precise potential -- in a system of $2\times10^6$ atoms at room temperature with periodic boundary conditions.
\Cref{fig:cp:speedup} shows that for large $x$ (near to 1) the speedup $s$ is limited by the required ACE computations.
However, a speedup of one or two orders of magnitude is possible when only a small fraction of ACE computations is required ($x$ near to 0).
In the latter case, the speedup is limited by the overhead caused mainly by the computation of descriptor (\cref{eq:cp:descriptor}) and switching parameter (\cref{eq:cp:lambda}).
Therefore, a fast descriptor is desirable as the speedup depends on both the descriptor and the averaging radius.

\begin{figure}
\includegraphics[width=3.37in]{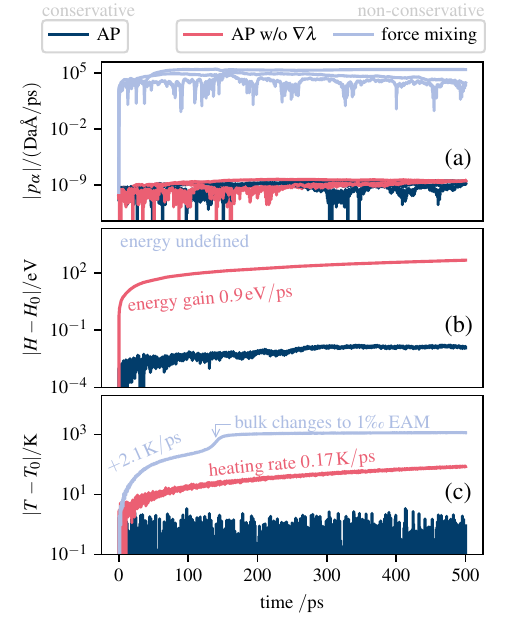}
\caption{\label{fig:cp:conservation}
a) Momentum, b) energy and c) temperature change in a \emph{NVE} simulation dependent on the combination method of EAM and ACE potential.
}
\end{figure}

The time integration, e.g., with the velocity-Verlet integrator~\cite{velocity_verlet}, of an atomic system without external forces described by a conservative interatomic potential results in a \emph{NVE} ensemble, i.e., number of particles, volume and total energy are conserved.
We assess this conservativity at room temperature for a system of 20100 W atoms with two surfaces, that is shown in Ref. \cite{sm}.
For non-conservative potentials (like LOTF potentials~\cite{aside_lotf}\nocite{csanyi.g.2004a,kermode.j.2015a,jinnouchi.r.2020a,jinnouchi2019,shapeev2020}) and force-mixing (cf. \cref{eq:cp:force:mix}; here calculated with ML-MIX~\cite{birks2025efficientaccuratespatialmixing,aside_mlmix}\nocite{birks_github,lammps}), the total energy is not conserved, which may lead to numerical instability or the system heating up like shown in \cref{fig:cp:conservation}d and Refs. \cite{arXiv241211569,energy_based_switching_qmmm,fu2023}.
Therefore, one should not neglect the gradient of the switching function, as the more precise forces (cf. \cref{fig:cp:ap:rmse}b) come at the cost of the conservativity of the potential.
Force-mixing has the further disadvantage that it violates momentum conservation as shown in \cref{fig:cp:conservation}a.
In contrast, the total momentum change is numerically zero for the conservative AP W potential, i.e., Newton's third law applies within numerical precision (cf. \cref{fig:cp:conservation}b).
Furthermore, the AP potential conserves the energy within the limit of the finite timestep of the time integrator (cf. Refs. \cite{sm,Skeel2005,zaverkin2021,fu2023,TOXVAERD1983214}).

We have shown that one can construct a conservative adaptive-precision potential with accurate forces based on the interpolation of potential energies with a switching parameter $\lambda_i$ that is calculated from a locally averaged descriptor $\pmb{d}_i$.
This new approach requires the usage of a differentiable descriptor $\pmb{d}_i(\{\pmb{r}^\text{d}\},\pmb{\beta})$, that only depends on atomic distances $\pmb{r}^\text{d}$ and and non-spatial constants $\pmb{\beta}$ like a nuclear charge number (cf. Ref. \cite{aside_zdetection}\nocite{sm}) as the usage of spatial reference positions changes the total momentum.
The differentiable CSP works as descriptor for solids, but one can introduce adaptive-precision also in other materials like fluids~\cite{Jesudason2007,doi:10.1021/acs.jctc.0c01112}.
In general, the descriptor needs to be small (cf. \cref{fig:cp:vis:scheme}a) for all atoms, which are well described by the fast surrogate model, so that the expensive precise model can be avoided.
Thereby, one can -- like demonstrated -- use exact potential energies and accurate forces for the atoms of interest, whereby the force accuracy improves when the averaging radius of $d_i$ is increased.
Therefore, force-based approaches, in which $\lambda_i$ interpolates directly the forces, are not required any more in order to obtain accurate forces on specific atoms.

Our approach can speed up simulations of microcanonical ensembles, in which the highest precision is required only for a subset of atoms, by a factor of up to 100.
Consequently, the usage of conservative AP potentials can highly contribute to a more energy-efficient computing.

\vspace{\baselineskip}
\emph{Acknowledgments} --- We thank A. Bochkarev for providing the preliminary ACE parametrization for tungsten.
D.I. provided formal analysis, investigation, methodology (equal), software, validation, visualization and writing of the original draft.
R.D. provided conceptualization (equal), methodology (equal), supervision (equal) and writing - review \& editing (equal).
G.S. provided conceptualization (equal), methodology (equal), supervision (equal) and writing - review \& editing (equal).

\vspace{\baselineskip}
\emph{Code availability} --- All simulations are done with LAMMPS~\cite{lammps}. Our modifications to the APIP package of LAMMPS are available at Ref. \footnote{\url{https://github.com/d-immel/lammps_ap/tree/apip_local_average}}.

\bibliography{refs.bib, asides.bib}

\section{End Matter}

\vspace{\baselineskip}
\emph{Appendix A: Force precision} ---
It is instructive to write the adaptive-precision force (\cref{eq:cp:force:hyb}) analogously to Refs. \cite{PhysRevB.102.024104,pace} as
\begin{equation}
\tag{A1}
\pmb{F}_i^\text{AP} = \sum_k \left( \pmb{f}_{ki}^\text{AP} - \pmb{f}_{ik}^\text{AP} \right)\,,
\label{eq:cp:force:from:contributions}
\end{equation}
since the force contributions $\pmb{f}_{ik}^\text{AP}$ for the adaptive-precision potential are thereby given as
\begin{equation}
\tag{A2}
\pmb{f}_{ki}^\text{AP} = \lambda_i \pmb{f}_{ki}^\text{(f)} + (1-\lambda_i) \pmb{f}_{ki}^\text{(p)} + \pmb{f}_{ki}^{(\lambda)}\,,
\label{eq:cp:force:contribution:total}
\end{equation}
where $\pmb{f}_{ki}^\text{(f)}$ and $\pmb{f}_{ki}^\text{(p)}$ can be calculated only from the fast and precise potential, respectively~\cite{sm}.
\Cref{eq:cp:energy:hyb,eq:cp:force:contribution:total}, i.e., the equations for the calculation of the adaptive-precision energy and force, have the same form, apart from the additional force contribution $\pmb{f}_{ki}^{(\lambda)}$ caused by the differentiation of the switching parameters.
Furthermore, \cref{eq:cp:force:from:contributions,eq:cp:force:contribution:total} show the dependence of $\pmb{F}_i^\text{AP}$ on the switching parameters $\lambda_k$ of all neighboring atoms $k$ within the cutoff.
Thus, although $\lambda_i=0$ implies a precise potential energy $E_i^\text{AP}$ (cf. \cref{eq:cp:energy:hyb}), it does not imply an accurate force $\pmb{F}_i^\text{AP}$.
Furthermore, it is important to note that although two models compute similar forces $\pmb{F}_i$, their force contributions $\pmb{f}_{ki}$ may not be correlated any more as shown in Fig. 1 of Ref. \cite{adaptive_precision_potentials}.
Thus, a slow spatial change of the switching parameter from 1 to 0 over few neighboring atoms is desirable to smoothly transfer the fast force model into the precise one.

\vspace{\baselineskip}
\emph{Appendix B: EAM-ACE W potential} ---
We construct a conservative adaptive-precision potential from the non-conservative adaptive-precision EAM-ACE W potential of Ref. \cite{paper_nanoindentation} in the following.

We adopt the atomic cluster expansion (ACE)~\cite{ace} W potential from Ref. \cite{paper_nanoindentation} as precise potential.
The force contributions $\pmb{f}^\text{ACE}$ (cf. \cref{eq:cp:force:contribution:total}) of an ACE potential are calculated in Refs. \cite{PhysRevB.102.024104,pace}.

Furthermore, we adopt the embedded atom model (EAM)~\cite{DAW1993251} W potential from Ref. \cite{paper_nanoindentation} as fast W potential.
The potential energy of an atom $i$ according to an EAM potential is
\begin{equation}
\tag{B1}
E_i^\text{EAM} = \xi\left(\zeta_i\right) + \frac{1}{2} \sum_{j\neq i} \Phi(r_{ij})\,,
\end{equation}
where $\xi$ is the embedding function, $\Phi$ a pair potential and $\zeta_i=\sum_{j\neq i}\zeta(r_{ij})$ the electron charge density at atom $i$.
One can calculate the force contributions $\pmb{f}^\text{EAM}_{ki}$ (cf. \cref{eq:cp:force:contribution:total}) analogously to Refs.~\cite{PhysRevB.102.024104,pace} as~\cite{sm}
\begin{equation}
\tag{B2}
\pmb{f}^\text{EAM}_{ki} =  \left( \left. \frac{\partial\xi}{\partial\zeta}\right|_{\zeta=\zeta_i}\left.\frac{\partial\zeta}{\partial r}\right|_{r=r_{ki}} + \frac{1}{2} \left.\frac{\partial\Phi}{\partial r}\right|_{r=r_{ki}}\right) \hat{\pmb{r}}_{ki}\,,
\label{eq:cp:force:contribution:eam}
\end{equation}
where $\hat{\pmb{r}}=\pmb{r}/r$ denotes a unit vector.

The centro-symmetry parameter (CSP)~\cite{csp} is used within a non-differentiable switching parameter in Ref. \cite{paper_nanoindentation}.
The CSP uses pairs of neighbors in exact opposite directions of atom $i$ to detect whether $i$ is near defects or surfaces.
In LAMMPS~\cite{lammps}, these neighbor pairs are recalculated every timestep for the calculation of $\text{CSP}^\text{dyn}_i$~\cite{larsen2020csp}.
Therefore, $\text{CSP}^\text{dyn}_i$ is not differentiable~\cite{larsen2020csp} and needs to be modified to be used for the calculation of a differentiable descriptor.
Thus, we detect these neighbor pairs initially and do not update them afterwards for the calculation of $\text{CSP}^\text{list}_i$, i.e.,
\begin{equation}
\tag{B3}
\text{CSP}_i^\text{list} = \sum_j \frac{\Theta^\text{in}_{ij}}{2} \left|\pmb{r}_{i,j} + \pmb{r}_{i,\Theta^\text{ngh}_{ij}}\right|^2\,,
\label{eq:cp:csp:list}
\end{equation}
where $\Theta^\text{in}_{ij}$ is 1 if $j$ is an atom in the pair-list of $i$ and 0 otherwise and $\Theta^\text{ngh}_{ij}$ denotes the atom, that was initially on the opposite side of $i$ than $j$.
Thereby, $\text{CSP}^\text{list}_i$ is differentiable and used as $g_i$ in the descriptor calculation (cf. \cref{eq:cp:descriptor}).
The CSP fluctuates around zero for atoms at a finite temperature in a the defect-free lattice, for which the fast EAM potential was optimized~\cite{paper_nanoindentation}.
Therefore, the reference descriptor $d^{(0)}$ (cf. \cref{eq:cp:descriptor}) is zero for this potential.
For simplicity, we use $w(r) = f^\text{trans}(r, r_\text{w}^\text{cut}-\SI{1}{\angstrom}, r_\text{w}^\text{cut})$ as weighting function for the averaging within the descriptor (cf. \cref{eq:cp:descriptor}).
Thereby, thermal fluctuations of atoms near the averaging cutoff $r_\text{w}^\text{cut}$ change the averaged descriptor smoothly rather than abruptly.

Finally, the switching parameter $\lambda_i$ is calculated according to \cref{eq:cp:lambda} from $d_i$.
$\text{CSP}_i \leq \SI{1.5}{\angstrom^2}$ is expected in body-centered cubic tungsten at room temperature due to thermal fluctuations of atoms~\cite{paper_nanoindentation}.
By using $d_\text{hi}=\SI{1.5}{\angstrom^2}$ (cf. \cref{eq:cp:lambda}) we ensure that only atoms with a higher than for thermal fluctuations expected CSP cause the evaluation of the precise ACE potential.
$d_\text{lo}$ (cf. \cref{eq:cp:lambda}) is set just high enough so that all atoms in defect-free bulk have $\lambda_i=1$, i.e., require only the fast EAM computation.
The force contribution $\pmb{f}_{ik}^{(\lambda)}$ from the differentiation of the switching parameter is~\cite{sm}
\begin{equation}
\tag{B4}
\begin{split}
\pmb{f}_{ik}^{(\lambda)} =& \Theta_{ik}^\text{in} \left(\sum_j \frac{E_j^{(\text{f})}-E_j^{(\text{p})}}{\overline{w}_j}\frac{\partial\lambda_j}{\partial d_j} w(r_{ji}) \right) 2 \left(\pmb{r}_{i,k} + \pmb{r}_{i,\Theta_{ik}^\text{ngh}}\right) \\
&+ \frac{E_i^\text{(f)}-E_i^\text{(p)}}{\overline{w}_i} \frac{\partial\lambda_i}{\partial d_i} \left(g_k-d_i\right) \left.\frac{\partial w}{\partial r}\right|_{r=r_{ik}} \hat{\pmb{r}}_{ik}\,.
\end{split}
\label{eq:cp:force:contribution:lambda}
\end{equation}

One can introduce atoms $i$ with a constant switching parameter as described in Ref. \cite{paper_nanoindentation}.
Thereby, one can enforce the fast or the precise calculation for a subset of atoms for a whole simulation.
\Cref{eq:cp:force:contribution:lambda} remains valid in this case, where $\frac{\partial\lambda_i}{\partial d_i}=0$
applies for the corresponding atoms as $\lambda_i$ is constant and not calculated from $d_i$ (cf. \cref{eq:cp:lambda}).

Atoms need a unique identifier due to the pair-list (cf. \cref{eq:cp:csp:list}).
Thus, this exemplary potential is not invariant under permutation of atoms any more.
Note that the general averaging approach (\cref{eq:cp:descriptor,eq:cp:lambda}) is invariant under permutation of atoms if $\pmb{g}_i$ is invariant under permutation of atom indices.

\onecolumngrid
\clearpage
\section{Supplemental material}
\setcounter{figure}{0}
\renewcommand{\thefigure}{S\arabic{figure}}

\subsection{Transition function}
As described in the manuscript, $f^{(\text{trans})}$ is taken from Eqs. (B2) and (B3) in Ref. \cite{adaptive_precision_potentials}, i.e.,
\begin{equation}
\tag{S1}
f^\text{(trans)} (x, x_\text{lo}, x_\text{hi}) =
\left\{
\begin{array}{ll}
1 & \text{for } x \leq x_\text{lo}\,,\\
0.5 + 0.9375 \frac{x_\text{hi}+x_\text{lo}-2x}{x_\text{hi} - x_\text{lo}} - 0.625 \left(\frac{x_\text{hi}+x_\text{lo}-2x}{x_\text{hi} - x_\text{lo}}\right)^3 +0.1875 \left(\frac{x_\text{hi}+x_\text{lo}-2x}{x_\text{hi} - x_\text{lo}}\right)^5 & \text{for } x_\text{lo} < x < x_\text{hi}\,,\\
0& \text{for } x_\text{hi} \leq x  \,,
\end{array}
\right.
\label{sm:eq:f:trans}
\end{equation}
as the first two derviatives of $f^{(\text{trans})}$ are smooth at $x_\text{lo}$ and $x_\text{hi}$.
$f^{(\text{trans})}$ is shown in Fig. S1.
$f^{(\text{trans})}$ is used in the weighting function $w(r) = f^{(\text{trans})}(r,r_\text{w}^\text{cut}-\SI{1}{\angstrom},r_\text{w}^\text{cut})$ to calculate the descriptor
\begin{equation}
\tag{S2}
\pmb{d}_i = \sum_k \pmb{g}_k(\{\pmb{r}^\text{d}\},\pmb{\beta}) \frac{w(r_{ik})} {\overline{w}_i(\{\pmb{r}^\text{d}\})} - \pmb{d}^{(0)}\,,
\label{sm:eq:overline:d:i}
\end{equation}
and to calculate the switching parameter
\begin{equation}
\tag{S3}
\lambda_i = f^{(\text{trans})}(d_i, d_\text{low}, d_\text{high})
\label{sm:eq:lambda}
\end{equation}
itself.
\begin{figure}[h]
\centering
\begin{tikzpicture}
\begin{axis}[
xlabel = $x$,
ylabel = $f^\text{trans}$,
every axis plot/.append style={ultra thick},
every tick/.style={thick},
axis line style = thick,
]
\addplot[color=fzjblue, domain=0:2, forget plot]{1};
\addplot[color=fzjblue, domain=2:4, samples=1000]{0.5+0.9375*(4+2-2*x)/(4-2)-0.625*((4+2-2*x)/(4-2))^3+0.1875*((4+2-2*x)/(4-2))^5};
\addplot[color=fzjblue, domain=4:7, forget plot]{0};
\addlegendentry{$f^\text{(trans)}(x,2,4)$};

\addplot[color=fzjred, domain=0:1, forget plot]{1};
\addplot[color=fzjred, domain=1:6, samples=1000]{0.5+0.9375*(6+1-2*x)/(6-1)-0.625*((6+1-2*x)/(6-1))^3+0.1875*((6+1-2*x)/(6-1))^5};
\addplot[color=fzjred, domain=6:7, forget plot]{0};
\addlegendentry{$f^\text{(trans)}(x,1,6)$};
\end{axis}
\end{tikzpicture}
\caption{
\label{sm:fig:f:trans}
$f^\text{(trans)}(x, x_\text{lo}, x_\text{hi})$ according to \cref{sm:eq:f:trans}.
}
\end{figure}
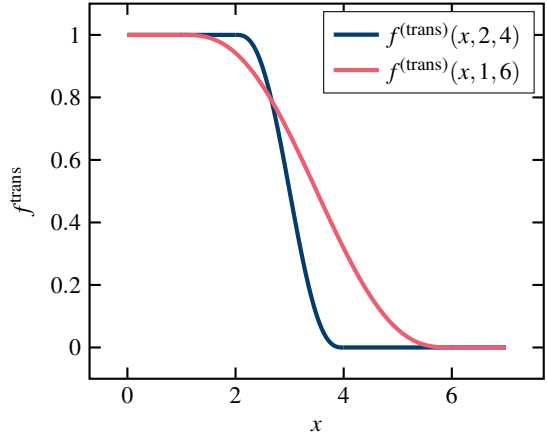

\subsection{Force calculation}
\label{sm:sec:force:calculation}
As stated in the manuscript, our adaptive-precision force is calculated as
\begin{equation}
\tag{S4}
\pmb{F}_i  = -\nabla_i \sum_k E_k = -\nabla_i \sum_k \left(\lambda_k E_k^\text{(f)} + (1-\lambda_k) {E}_k^\text{(p)} \right) = \sum_k \left(- \lambda_k \nabla_i E_k^\text{(f)} - (1 - \lambda_k) \nabla_i E_k^\text{(p)} + (\nabla_i\lambda_k) (E_k^\text{(p)} - E_k^\text{(f)})\right)\,.
\label{sm:eq:force:hyb:1}
\end{equation}
One can write the adaptive-precision force (\cref{sm:eq:force:hyb:1}) analogously to Refs. \cite{PhysRevB.102.024104,pace} as
\begin{equation}
\tag{S5}
\pmb{F}_i = \sum_k \left( \pmb{f}_{ki} - \pmb{f}_{ik} \right)\,,
\label{sm:eq:eamace:f:k:ap:sum}
\end{equation}
since the force contributions $\pmb{f}_{ik}$ for the adaptive-precision potential are thereby given as
\begin{equation}
\tag{S6}
\pmb{f}_{ki} = \lambda_i \pmb{f}_{ki}^\text{(f)} + (1-\lambda_i) \pmb{f}_{ki}^\text{(p)} + \pmb{f}_{ki}^{(\lambda)}\,,
\label{sm:eq:eamace:f:ki:ap}
\end{equation}
where the force contributions $\pmb{f}_{ki}^\text{(f)}$, $\pmb{f}_{ki}^\text{(p)}$ and $\pmb{f}_{ki}^{(\lambda)}$ correspond to fast potential, precise potential and switching function, respectively.
The force contributions of a fast embedded atom model (EAM)~\cite{DAW1993251}, a precise atomic cluster expansion (ACE)~\cite{ace} and of the switching parameter $\lambda$ are calculated in the following subsections.

\subsubsection{Embedded atom model}
\label{sm:sec:eam}
The energy of an atom $i$ described with EAM is
\begin{equation}
\tag{S7}
E_i^\text{EAM} = \xi\left(\zeta_i\right) + \frac{1}{2} \sum_{j\neq i} \Phi(r_{ij})\,,
\label{sm:eq:energy:eam}
\end{equation}
where $\xi$ is the embedding function, $\Phi$ a pair potential and the electron charge density $\zeta_i$ is given as
\begin{equation}
\tag{S8}
\zeta_i = \sum_{j\neq i}\zeta(r_{ij})\,,
\label{sm:eq:eam:zeta}
\end{equation}
where $\zeta(r_{ij})$ is a contribution to the electron charge density.
The contributions to \cref{sm:eq:force:hyb:1} caused by the differentiation of a fast EAM potential are
\begin{equation}
\tag{S9}
\pmb{F}_i^\text{AP,EAM} = \sum_k \left(- \lambda_k \nabla_i E_k^\text{EAM} \right)
= \sum_k \left(- \lambda_k \left( \left. \frac{\partial\xi}{\partial\zeta}\right|_{\zeta=\zeta_k}\sum_{j\neq k}\left.\frac{\partial\zeta}{\partial r}\right|_{r=r_{kj}} \hat{\pmb{r}}_{kj} \left(\delta_{ki}-\delta_{ji}\right) + \frac{1}{2} \sum_{j\neq k} \left.\frac{\partial\Phi}{\partial r}\right|_{r=r_{kj}} \hat{\pmb{r}}_{kj}\left(\delta_{ki} - \delta_{ji}\right) \right)\right)\,,
\label{sm:eq:eam:differentiation:0}
\end{equation}
where $\hat{\pmb{r}} = \pmb{r}/r$ is a unit vector.
Resolving the sums over the Kronecker deltas and rearranging the terms gives
\begin{equation}
\tag{S10}
\pmb{F}_i^\text{AP,EAM} = \sum_k \left( \lambda_i \left( \left. \frac{\partial\xi}{\partial\zeta}\right|_{\zeta=\zeta_i}\left.\frac{\partial\zeta}{\partial r}\right|_{r=r_{ki}} + \frac{1}{2} \left.\frac{\partial\Phi}{\partial r}\right|_{r=r_{ki}}\right) \hat{\pmb{r}}_{ki} - \lambda_k \left( \left. \frac{\partial\xi}{\partial\zeta}\right|_{\zeta=\zeta_k}\left.\frac{\partial\zeta}{\partial r}\right|_{r=r_{ik}} + \frac{1}{2} \left.\frac{\partial\Phi}{\partial r}\right|_{r=r_{ik}}\right) \hat{\pmb{r}}_{ik} \right) \,,
\label{sm:eq:eam:differentiation:1}
\end{equation}
where the introduction of the EAM pair force contribution
\begin{equation}
\tag{S11}
\pmb{f}^\text{EAM}_{ki} =  \left( \left. \frac{\partial\xi}{\partial\zeta}\right|_{\zeta=\zeta_i}\left.\frac{\partial\zeta}{\partial r}\right|_{r=r_{ki}} + \frac{1}{2} \left.\frac{\partial\Phi}{\partial r}\right|_{r=r_{ki}}\right) \hat{\pmb{r}}_{ki}
\label{sm:eq:f:ki:eam}
\end{equation}
allows to write the force as
\begin{equation}
\tag{S12}
\pmb{F}_i^\text{AP,EAM} = \sum_k \left(\lambda_i \pmb{f}^\text{EAM}_{ki} -\lambda_k \pmb{f}^\text{EAM}_{ik} \right) \,.
\label{sm:eq:eam:differentiation:2}
\end{equation}
The force on an atom $i$ according to a constant precision EAM potential can be directly obtained by assuming $\lambda=1$ for all atoms in \cref{sm:eq:eam:differentiation:0,sm:eq:eam:differentiation:2}, i.e.,
\begin{equation}
\tag{S13}
\pmb{F}_i^\text{EAM} = - \nabla_i \sum_k E_k^\text{EAM} = \left. \pmb{F}_i^\text{AP,EAM}\right|_{\lambda = 1} = \sum_k \left(\pmb{f}^\text{EAM}_{ki} - \pmb{f}^\text{EAM}_{ik} \right) \,,
\label{sm:eq:force:constant:precision:eam}
\end{equation}
where the EAM pair force contributions $\pmb{f}^\text{EAM}_{ki}$ (\cref{sm:eq:f:ki:eam}) remain valid, as they are independent of the switching parameter.

\subsubsection{Atomic cluster expansion}
\label{sm:sec:ace}
The force contributions $\pmb{f}^\text{ACE}_{ki}$ of an ACE~\cite{ace} potential are given in Eq. (42) of Ref. \cite{PhysRevB.102.024104} and Eq. (16) of Ref. \cite{pace}.
Comparing \cref{sm:eq:eam:differentiation:2,sm:eq:force:constant:precision:eam} shows that one only has to weight the constant-precision force contributions by a switching parameter to get an adaptive-precision force.
Thus, the contributions to the adaptive-precision force (cf. \cref{sm:eq:force:hyb:1}) by the differentiation of a precise ACE potential are
\begin{equation}
\tag{S14}
\pmb{F}_i^\text{AP,ACE} = - \sum_k (1 - \lambda_k) \nabla_i E_k^\text{ACE} = \sum_k \left((1 - \lambda_i) \pmb{f}^\text{ACE}_{ki} - (1 - \lambda_k) \pmb{f}^\text{ACE}_{ik} \right) \,.
\label{sm:eq:ace:force}
\end{equation}

\subsubsection{Switching parameter}
\label{sm:sec:lambda}
In general, the gradient of the switching parameter contributes to the adaptive-precision force (\cref{sm:eq:force:hyb:1}) in form of
\begin{equation}
\tag{S15}
\pmb{F}_i^{(\nabla\lambda)} = \sum_k \left(\nabla_i \lambda_k\right) \left(E_k^\text{(p)} - E_k^\text{(f)}\right)\,.
\label{sm:eq:force:lambda:0}
\end{equation}
For our definition of the switching parameter with a locally averaged detection mechanism (cf. \cref{sm:eq:overline:d:i,sm:eq:lambda}) follows
\begin{equation}
\tag{S16}
\pmb{F}_i^{(\nabla\lambda)} = \sum_k \left(E_k^\text{(p)} - E_k^\text{(f)}\right) \frac{\partial\lambda_k}{\partial d_k} \nabla_i d_k 
= \sum_k \tilde{E}_k \left(\sum_j \left(\nabla_i g_j\right) w(r_{kj}) + \sum_j \left(g_j -d_k\right) \left(\nabla_i w(r_{kj})\right) \right)\,,
\label{sm:eq:force:lambda:1}
\end{equation}
where the abbreviation
\begin{equation}
\tag{S17}
\tilde{E}_k = \frac{E_k^\text{(p)} - E_k^\text{(f)}}{\sum_j w(r_{kj})} \frac{\partial\lambda_k}{\partial d_k}
\label{sm:eq:e:tilde}
\end{equation}
is used.
The force $\pmb{F}_i^{(\nabla\lambda)}$ requires the calculation of $\nabla_i g_j$ and $\nabla_i w(r_{kj})$ (cf. \cref{sm:eq:force:lambda:1}).
As the former differentiation depends on the atomic quantities $g_j$, we start with the latter differentiation by calculating
\begin{equation}
\tag{S18}
\nabla_i w(r_{kj}) = \left.\frac{\partial w}{\partial r}\right|_{r=r_{kj}} \hat{\pmb{r}}_{kj} \left(\delta_{ik} - \delta_{ij}\right)
\label{sm:eq:nabla:w}
\end{equation}
and resolving the sums over the Kronecker deltas to
\begin{equation}
\tag{S19}
\pmb{F}_i^{(\nabla\lambda,\text{w})}= \sum_k \tilde{E}_k \sum_j \left(g_j -d_k\right) \nabla_i w(r_{kj})
= \sum_j \left(\tilde{E}_i \left(g_j -d_i\right) \left.\frac{\partial w}{\partial r}\right|_{r=r_{ij}} \hat{\pmb{r}}_{ij} - \tilde{E}_j \left(g_i -d_j\right) \left.\frac{\partial w}{\partial r}\right|_{r=r_{ji}} \hat{\pmb{r}}_{ji}\right)\,.
\label{sm:eq:force:lambda:2}
\end{equation}
With the force contribution
\begin{equation}
\tag{S20}
\pmb{f}_{ij}^{(\nabla\lambda,\text{w})}
= \frac{E_i^\text{(f)} - E_i^\text{(p)}}{\sum_j w(r_{ij})} \frac{\partial\lambda_i}{\partial d_i} \left(g_j -d_i\right) \left.\frac{\partial w}{\partial r}\right|_{r=r_{ij}} \hat{\pmb{r}}_{ij}
\label{sm:eq:force:contribution:nabla:w}
\end{equation}
follows
\begin{equation}
\tag{S21}
\pmb{F}_i^{(\nabla\lambda,\text{w})} = \sum_j \left(\pmb{f}_{ji}^{(\nabla\lambda,\text{w})} - \pmb{f}_{ij}^{(\nabla\lambda,\text{w})}\right)\,.
\label{sm:eq:force:nabla:w}
\end{equation}

\subsubsection{Centro-symmetry parameter}
\label{sm:sec:csp}
A modification of the centro-symmetry parameter\cite{csp} is used as atomic quantity $g_i$ for the calculation of the descriptor in the manuscript.
This differentiable CSP is given by
\begin{equation}
\tag{S22}
\text{CSP}_i^\text{list} = \sum_j \frac{\Theta^\text{in}_{ij}}{2} \left|\pmb{r}_{i,j} + \pmb{r}_{i,\Theta^\text{ngh}_{ij}}\right|^2\,,
\label{sm:eq:csp}
\end{equation}
where $\Theta^\text{in}_{ij}$ is 1 if $j$ is an atom in the pair-list of $i$ and 0 otherwise and $\Theta^\text{ngh}_{ij}$ denotes the atom, that was initially on the opposite side of $i$ than $j$.
For the $\nabla_ig_j$ term in \cref{sm:eq:force:lambda:1} follows with $\text{CSP}^\text{list}_j$ (\cref{sm:eq:csp}) as atomic quantity $g_j$
\begin{equation}
\tag{S23}
\nabla_i\text{CSP}_j^\text{list} = \sum_l \frac{\Theta^\text{in}_{jl}}{2} \nabla_i \sum_{\alpha=0}^2 \left(2 r_{j\alpha} - r_{l\alpha} - r_{\Theta^\text{ngh}_{jl}\alpha} \right)^2 = \sum_l \Theta^\text{in}_{jl} \left(\pmb{r}_{j,l} + \pmb{r}_{j,\Theta^\text{ngh}_{jl}}\right) \left(2\delta_{ij} - \delta_{il} -\delta_{i\Theta^\text{ngh}_{jl}}\right)\,,
\label{sm:eq:derivative:csp:0}
\end{equation}
where the contribution of the last two Kronecker deltas are identical as $\Theta^\text{in}_{jl} = \Theta^\text{in}_{j\Theta^\text{ngh}_{jl}}$ applies by definition of $\Theta^\text{ngh}_{jl}$ and the sum over all atoms $l$ is calculated.
Thus, the derivative of our CSP is given by
\begin{equation}
\tag{S24}
\nabla_i\text{CSP}_j^\text{list} = 2 \sum_l \Theta^\text{in}_{jl} \left(\pmb{r}_{j,l} + \pmb{r}_{j,\Theta^\text{ngh}_{jl}}\right) \left(\delta_{ij} - \delta_{il}\right)\,.
\label{sm:eq:derivative:csp:1}
\end{equation}
Therefore, one can calculate the force caused by the differentiation of the CSP in \cref{sm:eq:force:lambda:1} by resolving the sums over the Kronecker deltas, i.e.,
\begin{equation}
\tag{S25}
\pmb{F}_i^{(\nabla\lambda,\text{CSP})} = \sum_k \tilde{E}_k \sum_j \left(\nabla_i \text{CSP}^\text{list}_j\right) w(r_{kj})
= \sum_j 2 \left(\sum_k \tilde{E}_k w(r_{ki})\right) \Theta^\text{in}_{ij} \left(\pmb{r}_{i,j} + \pmb{r}_{i,\Theta^\text{ngh}_{ij}}\right) - \sum_j 2 \left(\sum_k \tilde{E}_k w(r_{kj})\right) \Theta^\text{in}_{ji} \left(\pmb{r}_{j,i} + \pmb{r}_{j,\Theta^\text{ngh}_{ji}}\right) \,,
\label{sm:eq:nabla:csp:2}
\end{equation}
and identifying the force contribution
\begin{equation}
\tag{S26}
\pmb{f}_{ij}^{(\nabla\lambda,\text{CSP})}
= \Theta^\text{in}_{ij} \left(\sum_k \frac{E_k^\text{(f)} - E_k^\text{(p)}}{\sum_j w(r_{kj})} \frac{\partial\lambda_k}{\partial d_k} w(r_{ki})\right) 2 \left(\pmb{r}_{i,j} + \pmb{r}_{i,\Theta^\text{ngh}_{ij}}\right)\,.
\label{sm:eq:nabla:csp:3}
\end{equation}
Thereby, the force can be simplified to
\begin{equation}
\tag{S27}
\pmb{F}_i^{(\nabla\lambda,\text{CSP})} = \sum_j \left( \pmb{f}_{ji}^{(\nabla\lambda,\text{CSP})} - \pmb{f}_{ij}^{(\nabla\lambda,\text{CSP})} \right)\,.
\label{sm:eq:nabla:csp:4}
\end{equation}
The force $\pmb{F}_i^{(\nabla\lambda)}$ (cf. \cref{sm:eq:force:lambda:0}) caused by the differentiation of the switching parameter requires the differentiation of the weighting function $w$ (cf. \cref{sm:eq:nabla:w,sm:eq:force:lambda:2,sm:eq:force:contribution:nabla:w,sm:eq:force:nabla:w}) and the atomic quantity $g_i$ (cf. \cref{sm:eq:derivative:csp:0,sm:eq:derivative:csp:1,sm:eq:nabla:csp:2,sm:eq:nabla:csp:3,sm:eq:nabla:csp:4}).
Thus, $\pmb{F}_i^{(\nabla\lambda)} = \pmb{F}_i^{(\nabla\lambda,\text{w})} + \pmb{F}_i^{(\nabla\lambda,\text{CSP})}$ is calculated from \cref{sm:eq:force:nabla:w,sm:eq:nabla:csp:4} to
\begin{equation}
\tag{S28}
\pmb{F}_i^{(\nabla\lambda)}
=  \sum_j \left( \pmb{f}_{ji}^{(\lambda)} - \pmb{f}_{ij}^{(\lambda)}\right)\,,
\end{equation}
where the force contribution $\pmb{f}_{ij}^{(\lambda)} = \pmb{f}_{ji}^{(\nabla\lambda,\text{CSP})} + \pmb{f}_{ji}^{(\nabla\lambda,\text{w})}$ follows from \cref{sm:eq:force:contribution:nabla:w,sm:eq:nabla:csp:3}, i.e.,
\begin{equation}
\tag{S29}
\pmb{f}_{ij}^{(\lambda)}
= \Theta^\text{in}_{ij} \left(\sum_k \frac{E_k^\text{(f)} - E_k^\text{(p)}}{\sum_j w(r_{kj})} \frac{\partial\lambda_k}{\partial d_k} w(r_{ki})\right) 2 \left(\pmb{r}_{i,j} + \pmb{r}_{i,\Theta^\text{ngh}_{ij}}\right) + \frac{E_i^\text{(f)} - E_i^\text{(p)}}{\sum_j w(r_{ij})} \frac{\partial\lambda_i}{\partial d_i} \left(g_j -d_i\right) \left.\frac{\partial w}{\partial r}\right|_{r=r_{ij}} \hat{\pmb{r}}_{ij}
\end{equation}
as stated in the manuscript.

\subsection{Energy conservation}
\begin{figure}[tb]
\begin{tikzpicture}[x=1in,y=1in]
\node[anchor=south west,inner sep=0] at (0,0) {\includegraphics[width=6.69in]{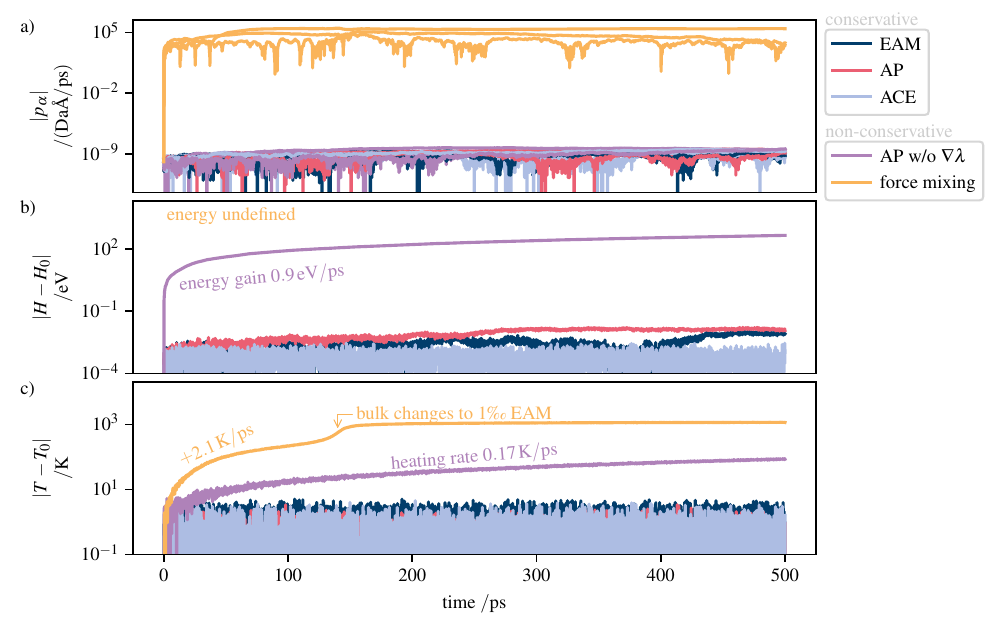}};
\node[anchor=south east,inner sep=0] at (6.2,0.4) {\includegraphics[width=0.35in]{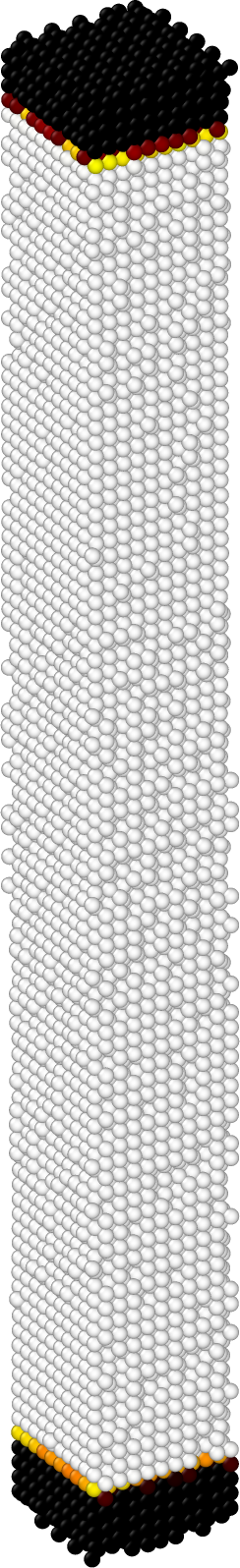}};
\draw[] (5.5,2.7) node[anchor=west] {d)};

\tikzmath{\cbw = 0.7;}
\node [coordinate] (cb_left) at (5.65,0.05) [] {};
\node [coordinate] (cb_w) at (\cbw,0.0) [] {};
\node [coordinate] (cb_right) at ($(cb_left) + (cb_w)$) [] {};
\node[anchor=west,inner sep=0] at (cb_left) {\frame{\includegraphics[width=\cbw in]{cb_hot.png}}};

\draw[] (cb_left) node[anchor=east] {ACE};
\draw[] (cb_right) node[anchor=west] {EAM};
\draw[] ($0.5*(cb_left)+0.5*(cb_right)$) node[anchor=south] {$\lambda_i\left(d_i\right)$};
\end{tikzpicture}
\caption{\label{sm:fig:conservation}
a) Total momentum, b) energy change and c) temperature change in a $\SI{500}{\pico\second}$ long \emph{NVE} simulation of the in d) shown system with the conservative adaptive-precision W potential with an averaging radius of $r_\text{w}^\text{cut}=\SI{16}{\angstrom}$.
The influence of non-conservatism is shown by neglecting the gradient of the switching function ($\nabla\lambda$) in the AP potential and by combining the EAM and ACE potential via force-mixing with ML-MIX~\cite{birks2025efficientaccuratespatialmixing}.
The conservative ACE and EAM potentials are shown as reference.
d) System of 20100 W atoms at room temperature with two surfaces visualized with OVITO~\cite{ovito}.
The atoms are colored according to the switching parameter $\lambda_i$ (cf. \cref{sm:eq:lambda}) of the conservative AP potential.
}
\end{figure}
\begin{figure}[tb]
\includegraphics[width=6.69in]{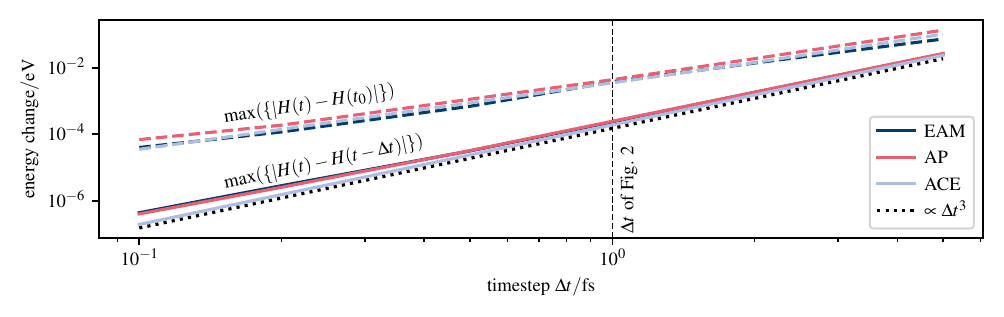}
\caption{\label{sm:fig:DeltaT}
Timestep dependence of the energy change between two consecutive timesteps and total energy change analyzed in a $\SI{10}{\pico\second}$ simulation starting after the simulations shown in Fig. S2.
An energy change proportional to $\Delta t^3$ is shown for comparison.
}
\end{figure}

The total energy $H$ of our conservative AP potential fluctuates systematically with a small amplitude on a short timescale and drifts randomly during the whole simulation to a negligible extend (cf. Fig. S2b).
Both, fluctuation and drift, are expected due to the finite timestep ($\SI{1}{\femto\second}$) of the time integrator~\cite{Skeel2005} (see also Refs. \cite{zaverkin2021,fu2023}).
The random energy fluctuation in the conservative AP simulation is in the same order as found in a reference simulation with the EAM potential, that is used as fast potential in the AP potential (cf. Fig. S2b).
To investigate the energy change further, we continued the simulation with the conservative potentials with different timesteps $\Delta t$ of the velocity-Verlet integrator.
Figure S3 shows a dependence of the energy change between two consecutive timesteps on $\Delta t^3$, i.e., the energy change vanishes with the timestep -- like expected from Ref. \cite{TOXVAERD1983214}.
Furthermore, the energy change between consecutive timesteps is similar for EAM and AP, while the energy change of ACE is smaller.
As the majority of the system uses the quickly computed EAM potential energy (cf. Fig. S2d), it is expected that the energy change of EAM dominates in AP compared to the ACE energy change.
Consequently, the usage of conservative AP potentials can highly contribute to a more energy-efficient computing.

\end{document}